\documentclass[twocolumn,showpacs,preprintnumbers,amsmath,amssymb]{revtex4}

\usepackage{graphicx}
\usepackage{amssymb}
\usepackage{amsfonts}
\usepackage{latexsym}
\input epsf
\input{epsf.sty}

\begin{document}


\title{Line nodes in the superconducting gap function of noncentrosymmetric CePt$_3$Si}

\author{K.~Izawa$^{1,2}$, Y.~Kasahara$^{1,3}$, Y.~Matsuda$^{1,3}$, K.~Behnia$^{1,4}$, T.~Yasuda$^5$, R.~Settai$^5$, and Y.~Onuki$^5$}
\affiliation{$^1$Institute for Solid State Physics, University of Tokyo, Kashiwanoha, Kashiwa, Chiba 277-8581, Japan}%
\affiliation{$^2$DRFMC/SPSMS/LCP, CEA-Grenoble 17, rue des Martyrs, 38054 Grenoble cedex 9 France}
\affiliation{$^3$Department of Physics, Kyoto University, Kyoto 606-8502, Japan}%
\affiliation{$^4$Laboratoire de Physique Quantique (CNRS), ESPCI, 10 Rue de Vauquelin, 75231 Paris, France}%
\affiliation{$^5$Graduate School of Science, Osaka University, Toyonaka, Osaka, 560-0043 Japan}%


\begin{abstract}

The superconducting gap structure of recently discovered heavy fermion CePt$_{3}$Si without spatial inversion symmetry was investigated by  thermal transport measurements down to 40~mK.   In zero field  a residual $T$-linear term was clearly resolved as $T\rightarrow 0$, with a  magnitude  in good agreement with the value expected for a residual normal fluid with a nodal gap structure, together with a $T^2$-dependence at high temperatures.   With an applied magnetic fields,  the thermal conductivity grows rapidly, in dramatic contrast to fully gapped superconductors,  and exhibits one-parameter scaling with $T/\sqrt{H}$.  These results place an important constraint on the order parameter symmetry,  that is CePt$_{3}$Si is most likely to have  line nodes.   

\end{abstract}


\maketitle @

	In heavy fermion (HF) compounds containing Ce, Pr and U atoms,  the strong Coulomb repulsion within the atomic $f$-shells leads to a notable many-body effect and  often gives rise to unconventional superconductivity,  in which the superconducting gap function has line or point nodes along certain directions in the Brillouin zone.   Since the gap structure is closely related to the pairing mechanism of the electrons,  the gap structure of HF superconductors is very important in understanding the physics of unconventional superconductivity in strongly correlated systems \cite{sig,thal}.
	 
	Very recently a new HF superconductor CePt$_{3}$Si with $T_{c}$=0.75~K has been discovered \cite{bauer}. CePt$_{3}$Si has aroused great interest because it possesses quite unique properties.   The most noticeable feature is the absence of spatial inversion symmetry,  in contrast to most unconventional superconductors.   It is well known that in the presence of strong spin-orbit interaction, the absence of inversion symmetry gives rise to a splitting of the two spin degenerate bands.  It has been shown that the lifted spin degeneracy strongly influences the pairing symmetry of Cooper pairs  \cite{bauer,sax,yip,gor,fri1,fri2,min,sam,sam2,cur}.  For example, Rashba-type spin-orbit interaction dramatically changes the paramagnetic effect, which breaks up Cooper pairs \cite{yip,gor,fri1,fri2,min}.   According to a subsequent band structure calculation,  the band splitting energy near the Fermi level in CePt$_3$Si is more than a thousand times larger than $k_{B}T_{c}$, indicating  strong spin-orbit coupling \cite{sam}.    The special interest in the symmetry of CePt$_{3}$Si has arisen with the observation that the upper critical field $H_{c2} \sim $ 4~T drastically exceeds the Pauli paramagnetic limit $H_{P} \sim1.3k_{B}T_{c}/\mu_{B} \sim$1~T \cite{bauer}.   Moreover, recent NMR measurements have reported no change of the Knight shift below $T_{c}$ for  {\boldmath $H$} $\parallel c$ \cite{kohara}.  These results indicate that the paramagnetic effect on  Cooper pairs  is absent or is strongly reduced in CePt$_{3}$Si.  Generally in the system without inversion symmetry  the gap function is a mixture of spin-singlet and -triplet channels in the presence of a finite spin orbit coupling strength.   The main channel of the gap function of CePt$_3$Si has been analyzed in the light of either spin-triplet  or  spin-singlet pairing \cite{bauer,fri1,fri2,sam,min,cur}.  Moreover,  peculiar properties,  including nonuniform helical superconducting phases \cite{heli} and the Zeeman-field-induced supercurrent \cite{fujimoto}, have been proposed in association with the lack of inversion symmetry. 
	
	Apart from exhibiting noncentrosymmetric superconductivity, CePt$_{3}$Si undergoes a superconducting transition after antiferromagnetic ordering with local moments, $\sim$0.2-0.3$\mu_B/$Ce,  occurs at $T_{N}$=2.2~K \cite{bauer}.     The ordered moments are coupled ferromagnetically in the tetragonal $ab$ plane and are stacked antiferromagnetically along the $c$-axis \cite{metoki,yogi}.    A similar coexistence has been reported in UPd$_{2}$Al$_{3}$ \cite{thal}.    In UPd$_{2}$Al$_{3}$,  strong coupling between  superconductivity and spin wave excitation has been reported \cite{sato} and the gap function $\Delta\propto \cos k_{z}c$ indicates  pairing between adjacent layers \cite{thal,watanabe}.  Therefore,  the relation between the superconductivity and local moment in CePt$_3$Si is an intriguing issue. 
	
	A major outstanding question about CePt$_{3}$Si is the nature of the superconducting gap structure, which is of primary importance for understanding the peculiar superconducting state associated with broken inversion symmetry and the coexistence of superconductivity and ordered moments.   The spin triplet  pairing state with point nodes, \mbox{\boldmath $d(k)$}=\mbox{\boldmath $\hat{x}$}$k_{y}$-\mbox{\boldmath $\hat{y}$}$k_{x}$,  has been proposed as a candidate for the gap function \cite{bauer,fri1}.   Recent NMR experiments have reported that the $T$-dependence of the relaxation rate $1/T_{1}$ below $T_{c}$ in CePt$_{3}$Si is different from $1/T_{1}$ for other HF superconductors, indicating a new class of  superconducting state with a point node gap function \cite{yogi}.    However, since the conclusion from NMR was drawn by the measurements  at high temperature and high field region in the vortex state,  a more detailed experimental investigation of the gap structure is required to shed light on the proposed novel superconducting states.   In this Letter, we report on thermal transport measurements down to 40~mK in order to clarify the gap structure of CePt$_3$Si.    We provide strong indication of line nodes in CePt$_3$Si,  in contrast to the previous studies.  
	
	Single crystals of CePt$_3$Si were grown by the Bridgeman method.  Clear de Haas-van Alphen oscillation was observed at low temperatures in sample from the same batch, confirming the high quality of the sample.  The thermal conductivity $\kappa$ was measured between 40~mK and 3~K  for a heat current  {\boldmath $q$} $\parallel$ $a$-axis of the tetragonal structure in {\boldmath $H$} $\parallel$ $b$-axis. 

\begin{figure}[b]
\begin{center}
\includegraphics[height=90mm]{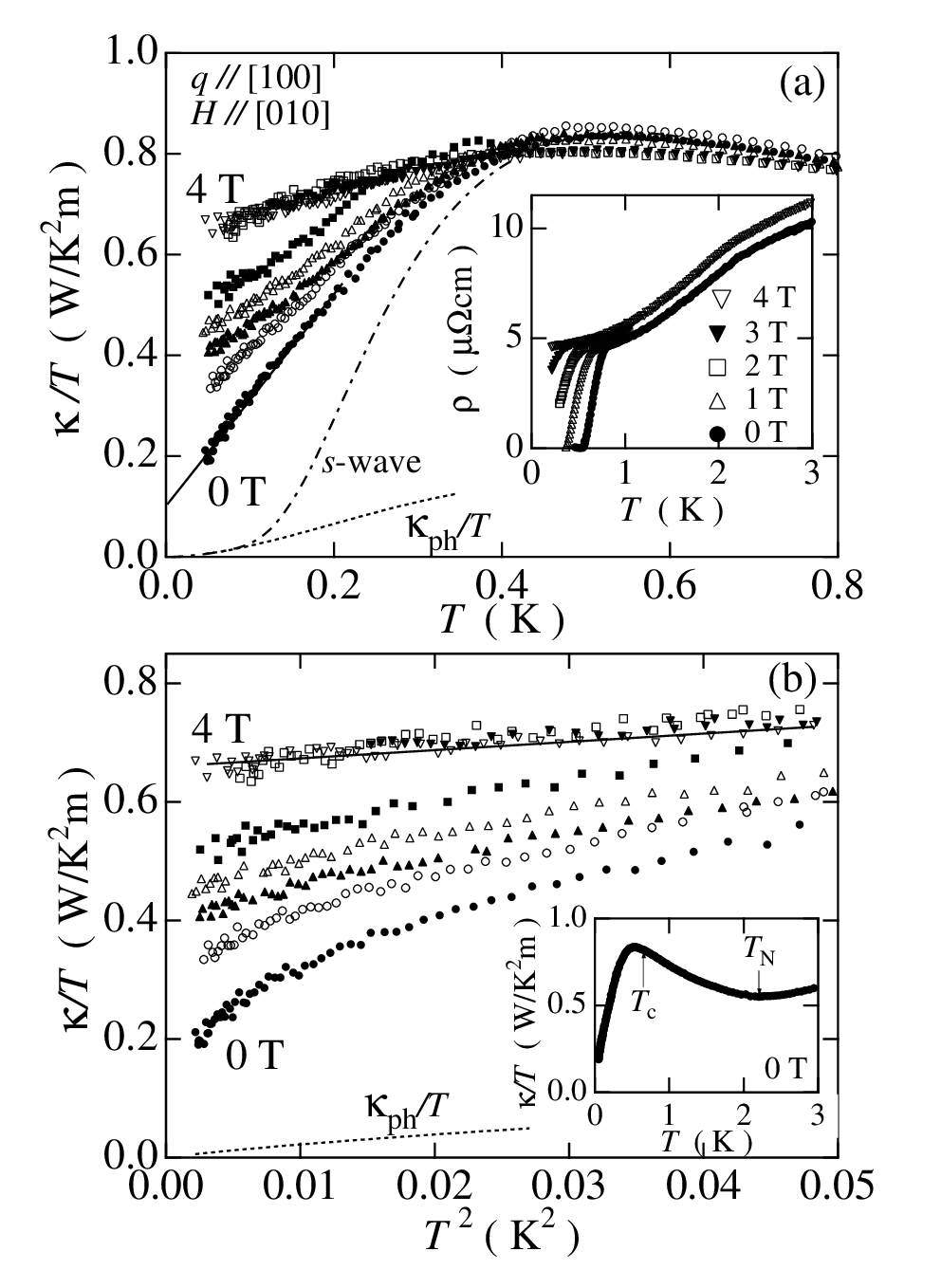}
\caption{(a) Main panel: $a$-axis thermal conductivity of CePt$_3$Si for {\boldmath $H$}$\parallel b$-axis, plotted as $\kappa/T$ vs. $T$ ($\bullet$ 0T, $\circ$ 0.2T, $\blacktriangle$ 0.5T, $\vartriangle$ 1T, $\blacksquare$ 1.5T, $\square$ 2T, $\blacktriangledown$ 3T, $\triangledown$ 4T).  The solid line is a linear fit  in zero field: $\kappa/T=A+BT$.  The dashed line is the contribution of phonons $\kappa_{ph}$.  The dash-dotted line is the thermal conductivity in $s$-wave superconductors.   Inset: $T$-dependence of $a$-axis resistivity for  {\boldmath $H$}$\parallel b$-axis.  (b)The same data at low temperatures plotted as $\kappa/T$ vs. $T^2$.  The solid line is a linear fit in the normal state: $\kappa/T=a+bT^2$.   The dashed line represents $\kappa_{ph}$.  Inset: $T$-dependence of $\kappa/T$ in zero field.   }
\end{center}
\end{figure}	
		
	 The insets of Fig.1(a) and (b) show the $a$ axis resistivity $\rho$ and $\kappa$.  Below $T_{N}$,  $\rho$ exhibits  Fermi liquid behavior, $\rho \propto T^2$.    At 50~mK, $H_{c2}$ determined by $\rho$ is slightly below 4~T.    Figure 1 and the inset of Fig.2(b) depict the $T$-dependence of $\kappa/T$.  In zero field, $\kappa/T$ exhibits a broad minimum at $T_N$,  shows no significant change at $T_{c}$ and decreases rapidly with decreasing $T$ after exhibiting a maximum just below $T_{c}$.  Similar behavior has been reported in UPt$_{3}$ \cite{sud}.   In order to use $\kappa$ as a probe of quasiparticle behavior, the phonon and antiferromagnetic magnon contributions must be extracted reliably.   The magnon contribution is negligibly small at $T<$~1~K, since the spin wave gap in CePt$_3$Si is reported to be larger than 15~K \cite{yogi}.  The simplest way to estimate the phonon conductivity  is to separate the phonon $\kappa_{ph}$ and electron $\kappa_{e}$ contributions in the normal state at a temperature sufficiently low that  $\kappa_{ph}$ has reached its well defined asymptotic $T^3$-dependent value given by
$\kappa_{ph}=\frac{1}{3}\beta \langle v_s \rangle T^3\ell_{ph}$, where $\beta$ is the phonon specific heat coefficient, $\langle v_s \rangle$ is the mean acoustic phonon velocity  and $\ell_{ph}$ is the phonon mean free path.  In Fig.~1(b), $\kappa/T$ is plotted as a function of $T^2$.  The solid line is a linear fit in the normal state,   $\kappa/T=a+bT^2$ with $a=$ 0.65~W/K$^{2}$m and $b=$ 1.4~W/K$^{4}$m.  Using $\langle v_s \rangle \sim$ 3000~m/sec and $\beta \simeq10.8$~J/K$^4$m$^3$ yields $\ell_{ph}\sim 200$~$\mu$m, which is comparable to the sample size.  Therefore the $T^3$-term in $\kappa$ mainly originates from phonons.  Note that the Lorentz number $L=\kappa \rho/T\simeq1.02L_0$ at $T\rightarrow 0$ in the normal state is very close to the Sommerfeld value $L_0=2.44\times10^{-8} $~$\Omega$ W/K.   The dashed lines in Figs.~1 and 2 show $\kappa_{ph}$.  Based on these estimations, we can conclude that heat transport well below $T_c$ is dominated by the electronic contribution: $\kappa\cong\kappa_{e}$.  

	 We first discuss the thermal transport in zero field.  As shown in Fig.~1(a), the $T$-dependence of $\kappa/T$ in CePt$_{3}$Si is markedly different from that in a fully gapped $s$-wave superconductor, in which $\kappa/T$ exhibits exponential behavior \cite{bard}.  As shown by the straight line in Fig.~1(a), $\kappa/T$ is well fitted to the data by $\kappa/T=A+BT$ in a wide $T$-range above 40~mK.  
	 
	  The presence of a residual term at $T\rightarrow 0$  in $\kappa/T$ is clearly resolved.   The residual term indicates the existence of a residual normal fluid, which is expected for an unconventional superconductor with nodes in the energy gap.   This residual normal fluid is a consequence of impurity scattering, even for low concentrations of nonmagnetic impurities \cite{lee,graf}.  It has been shown that for an order parameter with line nodes,  the quasiparticle thermal conductivity has components that are universal in the limit $T\rightarrow 0$ and of the form,
\begin{equation}
\frac{\kappa_e}{T}=\frac{\kappa_{00}}{T} \left( 1+ O\left[ \frac{T^2}{\gamma_0^2} \right] \right).
\end{equation}	 
in the range $k_{B}T<\gamma_{0}$,  where $\gamma_{0}$ is the impurity bandwidth \cite{lee,graf,maki}.  $\kappa_{00}/T$ is the residual term and the second $T^{2}$-term is a finite $T$ correction, which strongly depends on the impurity scattering phase shift.  We did not observe the $T^{2}$-term in $\kappa/T$ in our $T$ range.  This may be due to the fact that  $\gamma_{0}$ is comparable to or less than our lowest temperature.   In this case, the residual term extrapolated from high temperature regime, assuming  $\kappa/T=A+BT$,  gives a lower limit of $\kappa_{00}/T$, while the value at $\kappa/T$ at 40~mK gives the higher limit:  0.1W/K$^{2}$m$<\kappa_{00}/T<$0.2 W/K$^{2}$m.  The magnitude of the residual  term $\kappa_{00}/T$ for unitary scattering with small $\gamma_{0}$ is given by,
\begin{equation}
\frac{\kappa_{00}}{T}= \left(\frac{4}{\pi}\frac{\hbar\Gamma}{\Delta_{0}}\frac{1}{\mu}\right) \frac{\kappa_n}{T}.
\end{equation}	 
Here $\kappa_n$ is the normal state thermal conductivity, $\Delta_{0}$ is the maximum of the energy gap, $\Gamma$ ($\gamma_{0}\simeq 0.61 \sqrt{\hbar\Gamma\Delta_{0}}$) is the impurity scattering rate and  $\mu=1/\Delta_{0}\mid d\Delta(\phi)/d\phi \mid_{nodes}$ is the slope parameter of the gap function at the node.   This universal limit is the result of a compensation between the growth in the normal fluid density with increasing impurity concentration and the concomitant reduction in the mean free path of the quasiparticles \cite{lee,graf,maki}.   The universal thermal conductivity has been reported for high-$T_c$ cuprates \cite{tail} and Sr$_{2}$RuO$_{4}$ \cite{suzuki}.   Using $\hbar\Gamma/\Delta_{0} \simeq  \pi \xi/2\ell_e \sim$ 0.16 from de Haas-van Alphen\cite{dHvA} and $H_{c2}$ measurements, $\kappa_{00}/T$ is estimated to be $\sim$0.09W/K$^{2}$m, where $\xi$ is the coherence length and $\ell_e$ is the mean free path.   This value is close to the observed $\kappa_{00}/T\sim$ 0.1~W/K$^{2}$m.  Thus the residual conductivity is quantitatively consistent with a gap function with line nodes.

	The temperature dependence of $\kappa/T$ is intriguing.  Instead of being $T^2$ (as predicted by theory), it is $T$-linear.  Long ago, it has been argued that in the presence of line nodes in the gap function, the linear energy dependence of the density of states (DOS) leads to a $T$-linear $\kappa/T$ \cite{miyake}.  Experimentally, in several superconductors believed to host line nodes, such as CeCoIn$_{5}$ \cite{tana}, CeIrIn$_{5}$ \cite{mov} and Sr$_{2}$RuO$_{4}$ \cite{suzuki,izawa1}, $\kappa/T$ is $T$-linear in a wide temperature range below $T_{c}$. 

\begin{figure}[b]
\begin{center}
\includegraphics[height=90mm]{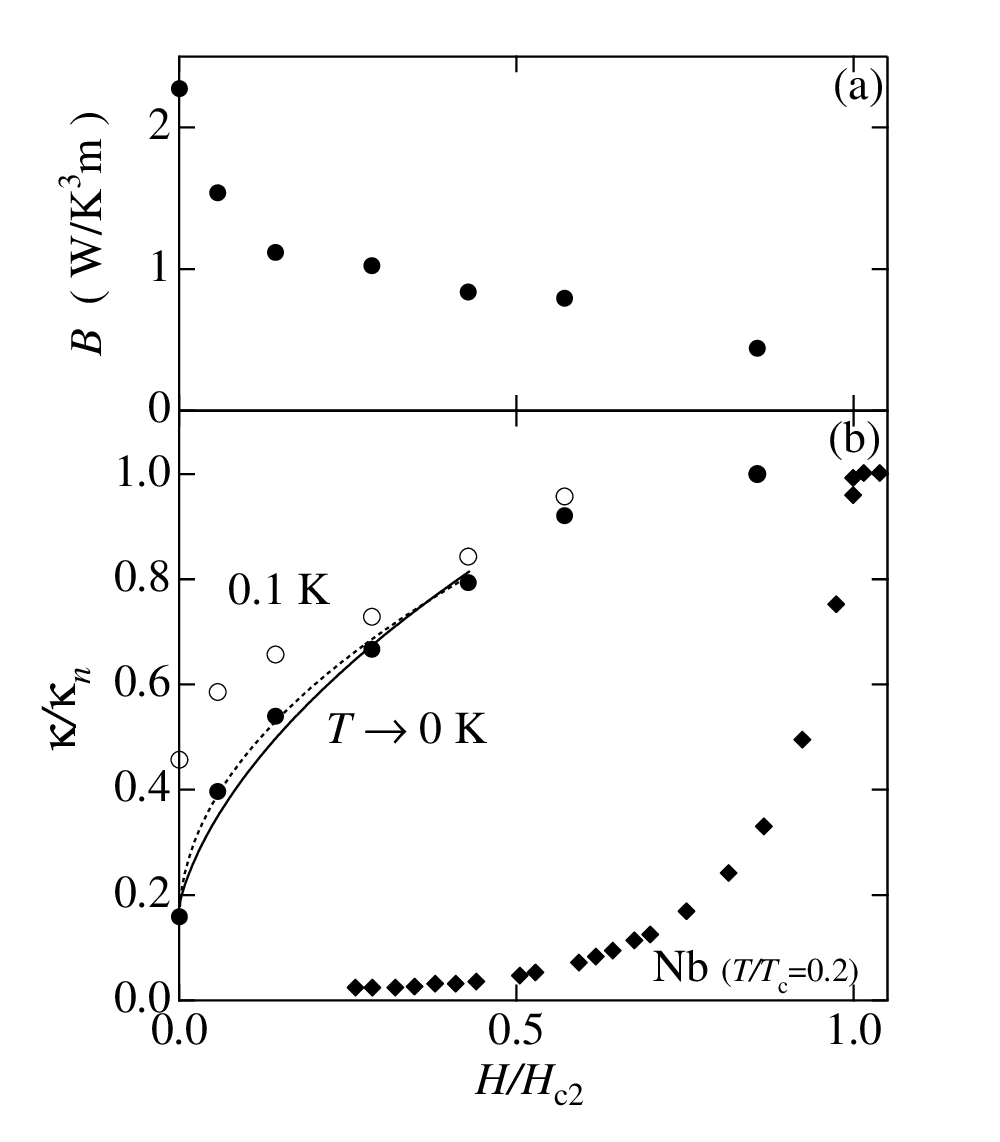}
\caption{(a)$H$-dependence of the $T$-linear coefficient of $\kappa/T=A+BT$, (b)$H$-dependence of $\kappa$,  normalized by the normal state value $\kappa_n$, as a function of $H/H_{c2}$ at $T=$100~mK and at  $T\rightarrow 0$.  We used $H_{c2}$= 3.6~T.   The dashed line represents $\kappa/\kappa_n=0.15+0.54H^{0.49}$.  The solid line is the result of fitting obtained by using Eq.~(3).  For comparison, data for the typical $s$-wave superconductor Nb at 2~K is shown ($\blacklozenge$). For details, see the text.   }
\end{center}
\end{figure}
	
	Strong indication for the presence of line nodes is provided by the $H$-dependence of $\kappa$.  Figure 2(b) depicts the $H$-dependence of $\kappa$ normalized by the normal state value $\kappa_n$ at 100~mK and at $T=0$ ($\kappa=\kappa_e$) obtained by the simple  extrapolation to $T\rightarrow0$.  For comparison, we show in Fig.~2(b) the thermal conductivity of the typical $s$-wave superconductor Nb at 2~K \cite{boa}.  The $H$-dependence of $\kappa/T$ in CePt$_{3}$Si  is in dramatic contrast to that in Nb.   In an $s$-wave superconductor, the only quasiparticle states present at $T \ll T_c$ are those associated with vortices.  When vortices are far apart, these states are bound to the vortex core and are therefore localized and unable to transport heat; the conduction shows an exponential behavior with a much slower growth with $H$ at $H \ll H_{c2}$.  Therefore  the large amount of delocalized quasiparticles throughout the vortex state of CePt$_3$Si is immediately apparent.  The $H$-dependence of $\kappa/\kappa_n$ at $T\rightarrow0$ is well fitted by $\kappa/\kappa_n=0.15+0.54H^{0.49}$ as shown by the dotted line in Fig.~2(b), indicating that the power law on $H$ is very close to 1/2.  In Fig.2(a), the $H$-dependence of the $T$-linear coefficient of $\kappa/T(=A+BT)$ is shown.  $B$ decreases faster than $\sqrt{H}$.  This may be important for understanding the origin of $T$-linear dependence of $\kappa/T$.
	
	   The understanding of the heat transport of superconductors with nodes has progressed considerably during past few years \cite{vekhter,kubert,hussey}.  In contrast to classical superconductors, the heat transport in nodal superconductors is dominated by contributions from delocalized quasiparticle states outside vortex cores.  The most remarkable effect on the thermal transport is the Doppler shift of the energy of quasiparticles with momentum {\boldmath $p$} in a circulating supercurrent flow $\mbox{\boldmath $v$}_s$  ($E(\mbox{\boldmath $p$})\rightarrow E(\mbox{\boldmath $p$})-\mbox{\boldmath $v$}_s \cdot \mbox{\boldmath $p$}$) \cite{volovik}.  This effect (Volovik effect) becomes significant at positions where the local energy gap becomes smaller than the Doppler shift term ($\Delta < \mbox{\boldmath $v$}_s \cdot \mbox{\boldmath $p$}$), which can occur in superconductors with nodes.  In the presence of line nodes, where the DOS of electrons $N(E)$ has a linear energy dependence ($N(E)\propto E$), $N(H)$ increases in proportion to $\sqrt{H}$.  According to  K\"{u}bert and Hirschfeld (KH),  the thermal conductivity at $T=0$ in superconductors with line nodes increases as	   
\begin{equation}
\kappa(0,H)=\kappa_{00} \frac{p}{p\sqrt{1+p^{2}}-\sinh^{-1}p}
\end{equation}
with $p \equiv \sqrt{8\hbar \Gamma H_{c2}/ \pi^{2} \Delta_0 H}$ for unitary limit scattering.  The solid line in Fig.~2 indicates the field dependence of $\kappa/T$ calculated from Eq.~(3)  using $\hbar \Gamma/\Delta_0=$ 0.10.  The $H$-dependence is quite well reproduced.  We stress here that this $\hbar \Gamma/\Delta_0$ value is close to that used for estimating the universal thermal conductivity, indicating the consistency between the two different experiments.

	To obtain further insight into the thermal conductivity in a magnetic field, we examined the scaling relation of the quasiparticle thermal conductivity with respect to $T$ and $H$ in accordance with Ref.\cite{kubert}.   At finite temperatures, the number of thermally excited qusiparticles exceeds that of Doppler shifted quasiparticles.  This crossover field is roughly estimated by $\sqrt{H/H_{c2}}\sim k_{B}T/\Delta_{0}$.  A scaling relation of the single variable $x=t/\sqrt{h}$ with $t=T/T_{c}$ and $h=H/H_{c2}$ is derived as
\begin{equation}
\kappa_e(T,H)/\kappa_e(T,0)=F(x),  
\end{equation}	
where $F(x)$ is a scaling function. Theoretically,  Eqs.(3) and (4) are expected to be valid only in the $H \ll H_{c2}$  regime.  As seen in Fig.3, the experimental data gives a satisfactory collapse for fields up to 1T (~0.3$H_{c2}(0)$) and for the $H$=1.5T data a deviation  from this collapse appears \cite{upt3}.  We here obtained $\kappa_e$ by subtracting $\kappa_{ph}$ from $\kappa$ at finite temperatures.   Since no scaling is expected for a point node, the present $T/\sqrt{H}$ scaling again strongly indicates the presence of line nodes.  We also note that the scaling behavior implies that scattering in CePt$_3$Si is in the unitary limit.

\begin{figure}[t]
\begin{center}
\includegraphics[height=80mm]{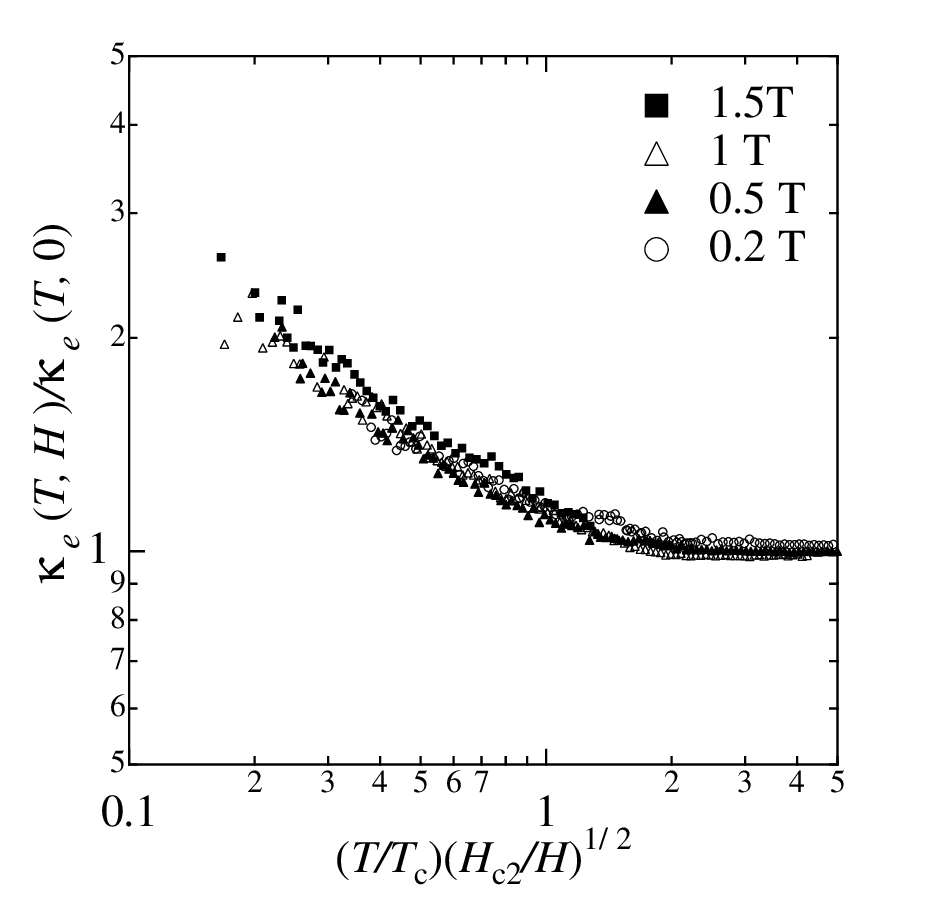}
\caption{Scaling plot $F\equiv\kappa_{e}(T,H)/\kappa_{e}(T,0)$ vs. $x=(T/T_c)/\sqrt{H/H_{c2}}$.  Note that the data collapse into the same curve. }
\end{center}
\end{figure}

	Summarizing the salient features of the superconductivity of CePt$_3$Si:  (1) a residual term in $\kappa(T,0)/T$ as $T\rightarrow 0$ whose magnitude is in good agreement with the universal conductivity, (2) $T$-linear dependence of $\kappa(T,0)/T$ at high temperatures,  (3) $H$-dependence well fitted by the theory of Doppler shifted quasiparticles and (4) scaling relations in the variable $T/\sqrt{H}$.   The combination of all these results lead us to conclude that the gap function of CePt$_3$Si  has line nodes.   Recently it has been reported that the $T$- and $H$-dependencies of the thermal conductivity in  $s$-wave superconductors with multigaps, such as MgB$_2$ and NbSe$_2$, exhibit a strong deviation from those in  $s$-wave supercnductors with a single gap \cite{nbse}.  However it is very unlikely that such a multigap scenario provide an explanation for the thermal conductivity in CePt$_3$Si.   Indeed, the presence of a finite residual term in the zero-field temperature dependence of $\kappa/T$ in temperatures as low as 40~mK puts an upper limit on the magnitude of any hypothetical second gap.  Moreover the scaling observed here is not expected in the multigap scenario.  At this stage, it is safe to conclude that we are in presence of a gap function with line nodes.  	
	
	The possibility of the line node in CePt$_3$Si was first discussed in Ref.\cite{sam}.  Recently Sergienko and Curnoe showed that the gap function in superconductors with lifted spin degeneracy can be expressed as  $\Delta(\mbox{\boldmath $k$})=\chi(\mbox{\boldmath $k$})t(\mbox{\boldmath $k$})$, where $\chi(\mbox{\boldmath $k$})$ is an even auxiliary function that transforms according to the irreducible representation and $t(\mbox{\boldmath $k$})$ is a model dependent phase factor \cite{cur}.  The thermodynamic properties are determined by the orbital part of $\chi(\mbox{\boldmath $k$})$.  The presence of line nodes is incompatible with the recent theoretical consideration of spin triplet gap function with \mbox{\boldmath $d(k)$}=\mbox{\boldmath $\hat{x}$}$k_{y}$-\mbox{\boldmath $\hat{y}$}$k_{x}$ (which corresponds to $\chi(\mbox{\boldmath $k$})=k_{x}^{2}+k_{y}^{2}$) and  NMR measurements, both of which suggest a point-node gap function.  The presence of antiferromagnetism in CePt$_3$Si may favour superconducting states for which $\chi$(\mbox{\boldmath $k+Q$})$ = - \chi$(\mbox{\boldmath $k$}) with  \mbox{\boldmath $Q$}$=(0,0,\pi/c)$ is satisfied \cite{sig}.    The functions with this property in a $C_{4v}$ crystal can be found in $A_1-$ and $E-$ representations   (Table I of Ref.\cite{cur}).  For $A_1$, it is $\cos(k_z c)$, which has line nodes at $k_z =\pm \pi/2c$. There are three possible states for $E$ representation, which all have line nodes at $k_z=0$ and obey $\chi$(\mbox{\boldmath $k+Q$})$ = - \chi$(\mbox{\boldmath $k$}): (a)$\sin k_z (\sin k_x + i \sin k_y)$, (b) $\sin k_z \sin k_x$ (or $\sin k_z \sin k_y$  another domain), (c) $\sin k_z (\sin k_x + \sin k_y)$.  


	In conclusion, we have investigated low energy quasiparticle excitation in CePt$_3$Si by  thermal conductivity measurements.   A residual linear term $\kappa_{00}/T$ whose magnitude is in good agreement with the universal  conductivity  is clearly observed, together with a quadratic term at high temperatures.    The field dependence is in dramatic contrast with that of fully gapped superconductors and exhibits one-parameter scaling with $T/\sqrt{H}$.  All of these results provide strong indication of line nodes in the gap function, in contrast to  recent theoretical and experimental studies.  Our results place strong constraints on models that attempt to explain the uncommon superconductivity without space inversion symmetry.
			
	We thank D.~Agterberg, S.H.~Curnoe, S.~Fujimoto, N.E.~Hussey, H.~Ikeda, H.~Kusunose,  K.~Maki, I.A.~Sergienko, H.~Shimahara, M.A.~Tanatar, P.~Thalmeier, and I.~Vekhter for valuable discussions.


\begin{thebibliography}{99}
\bibitem{upt3}Similar scaling relation has been reported for UPt$_3$ with line nodes, but the universal conductivity was not observed \cite{sud}.
\end{thebibliography}
\end{document}